\documentclass[preprint,showpacs,preprintnumbers,amsmath,amssymb]{revtex4}
\usepackage{epsf}
\def\E{{\mathcal E}}
\def\Z{{\mathbb Z}}
\begin{document}


\title{Layering in Crumpled Sheets}
\author{David Aristoff}
\affiliation{}
\author{Charles Radin}
\affiliation{Mathematics Department, University of Texas, Austin, TX 78712} 


\begin{abstract}
We introduce a toy model of crumpled sheets. 
We use simulation to show there is a first order phase transition in 
the model, from a disordered dilute phase to a mixture with a layered phase.
\end{abstract}


\maketitle

\section{Introduction}

When a sheet of stiff paper is crumpled into a compact ball, creases
and folds appear. This storage of energy, especially in the
irreversibly distorted creases, has been widely studied, for instance
in [1--4].  Our interest here is in geometric changes associated with
the (reversible) folds, which is less well understood. See for
instance [5,6].

Consider the densest possible state of the material, in which the
sheet is carefully folded into a compact stack of parallel
leaves. (There needs to be a significant cost for bending
the material or else dense packings will usually be more
complicated.)  Imagine the process of
compactifying the sheet within a contracting sphere, from a typical
initial state of low volume fraction near 0 to a typical state of high
volume fraction near 1. How would such a process 
proceed? For a material in thermal equilibrium there is a pair of
first order phase transitions associated with quasistatic
compactification; low density configurations are of random character
(fluid), while high density configurations are ordered (solid), and
isothermal compactification would progress between these extremes via
freezing and then melting transitions, separated by phase coexistence in
which the material consists of macroscopic portions of each phase.
Our computations, in a toy model, suggest that compactification of a
stiff sheet undergoes a similar path, at least from its low volume fraction
disordered state to a reorganization of the state into an inhomogeneous
intermediary (part random and part folded), as is typical of freezing.

Our toy model is not of a thin, stiff sheet in three dimensions but of a
thin, stiff wire loop in two dimensions, which we expect to behave
similarly [7]. (Confined wire in three dimensions may well behave
differently; see for instance [8].)
In the model the wire is restricted to 
a triangular lattice, so a configuration in the model is a
self-avoiding closed walk (loop) on the lattice, with energies
associated to bends in the walk to penalize the bends.

\begin{figure}\label{fig1}
\centerline{\epsfxsize=7truein \epsfbox{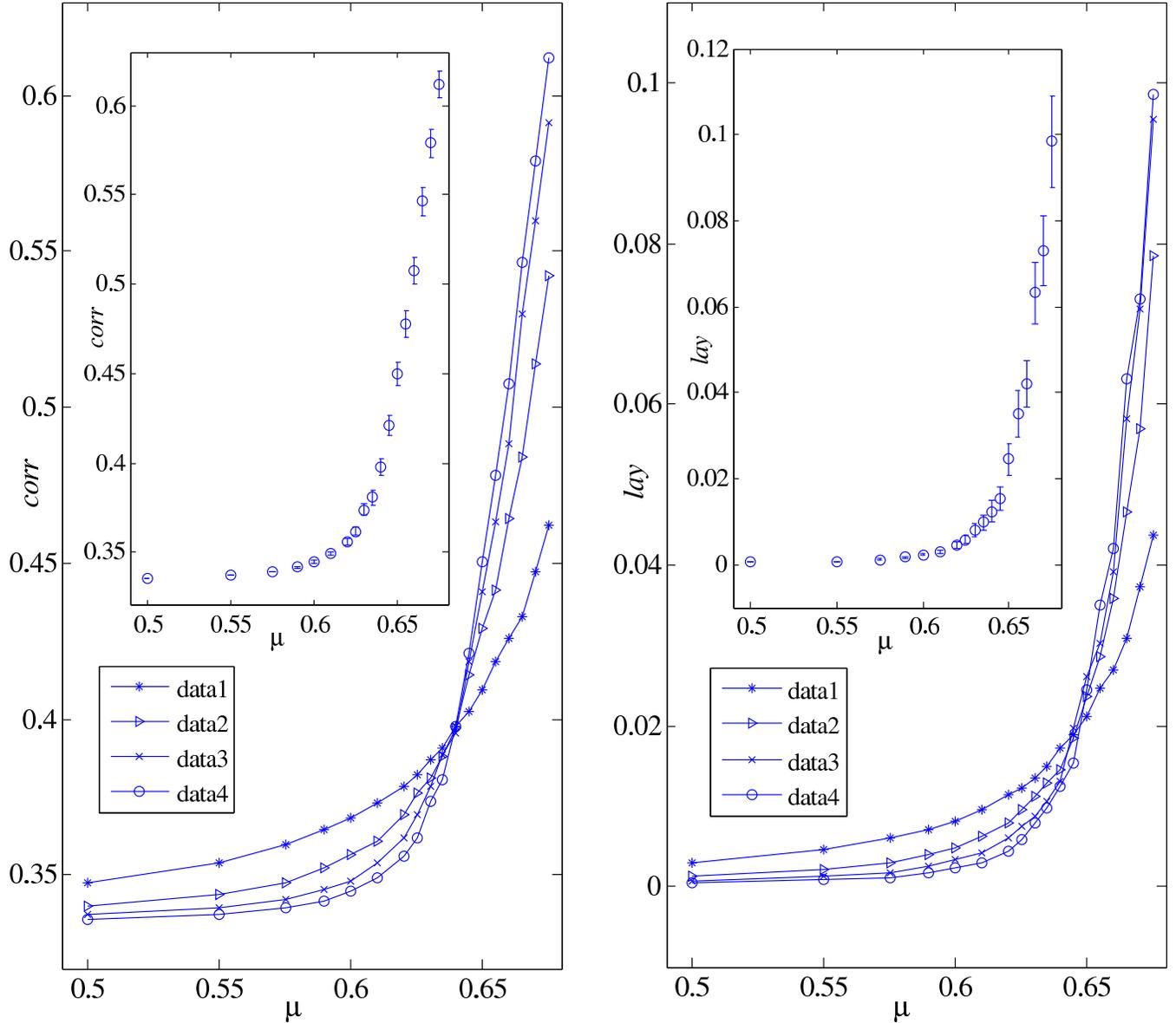}}
\caption{a) Correlation vs. mu,
for volumes $40^2$ (data1) through $100^2$ (data4);
b) Layer size vs. mu, for volumes $40^2$ (data1) through $100^2$ (data4).
Error bars in both insets represent 95$\%$ confidence intervals for
volume $100^2$, and for
low $\mu$ are
smaller than the data circles.}
\end{figure}

\begin{figure}\label{fig2}
\centerline{\epsfxsize=7truein \epsfbox{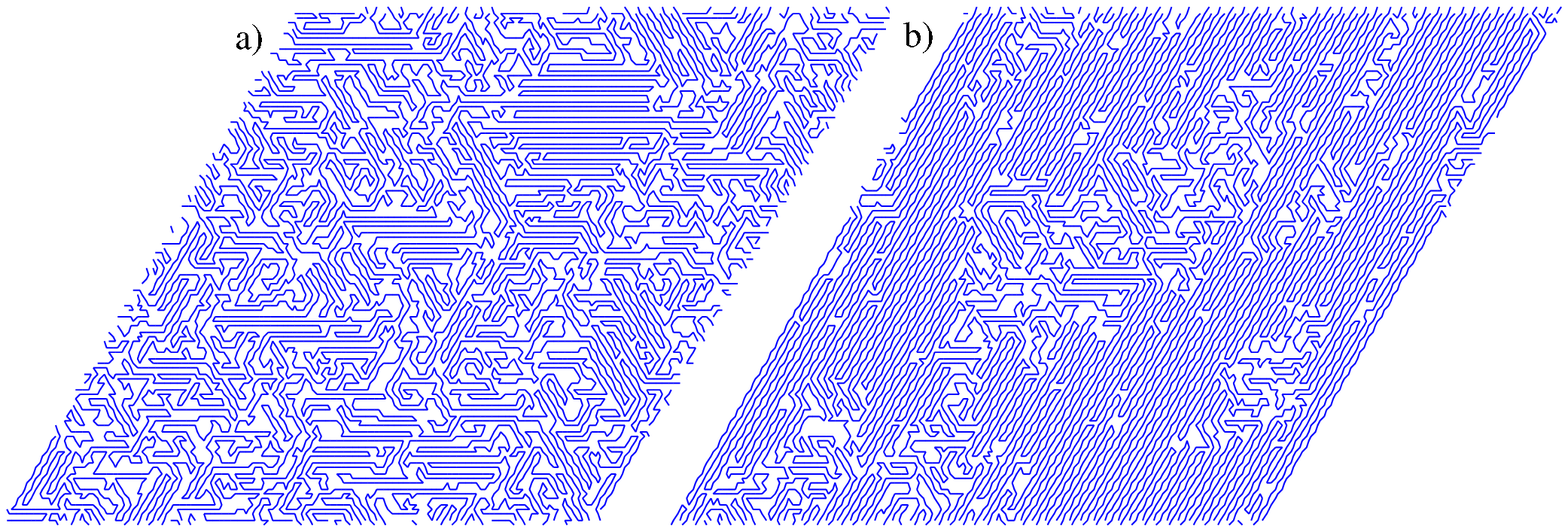}}
\caption{Snapshots of a loop in available volume $100^2$ in
  equilibrium at $\mu=0.6$ in a), and $\mu=0.67$ in b).}
\end{figure}

\section{The Model and Results}

For a fixed integer $n$ consider the triangular lattice $L = \{(a+b/2,
b\sqrt{3}/2) : (a,b) \in (\Z/n\Z)^2\}$ with periodic boundary
conditions. Note that this space is homogeneous and isotropic.

Let $\cal A$ be the set of oriented self-avoiding closed walks on $L$,
its elements being called {\it configurations}.
If a walk $C \in \cal A$ changes direction at a vertex, we say there 
is a bend there. If it changes by $\pm \pi/3$, we call the bend large; 
if it changes by $\pm \pi/6$ we call it small. 
We define $B_s(C)$ as the number of small bends in $C$ and 
$B_{\ell}(C)$ as the number of large bends in $C$. 
Assigning the energy $e_s$ to each small bend and $e_{\ell}$ to each large bend,
we define the total energy of configuration $C$ by:
$$\E(C)= B_s(C) e_s  + B_{\ell}(C) e_{\ell},
\eqno{(1)}$$ and denote the model with energy parameters $e_s$ and
$e_{\ell}$ by $e_s:e_{\ell}$.

Because it is harder to simulate our model at fixed density and/or
fixed energy, we use conjugate variables $\beta$ for energy $\E$ and
$\mu$ for particle (edge) number $N$, assigning the probability $m_\mu(C)$
for any $C \in \cal A$:
$$m_\mu(C) = \frac{e^{-\beta [\E(C)-\mu N(C)]}}{Z},\eqno{(2)}$$
where $Z$ is the normalization. (We suppress dependence on the system 
size.) We take $\beta = 1$ without loss of generality, so that the variable parameters 
in our model are $\mu$ and the energies of the two bending angles, 
$e_{\ell}$ and $e_s$. We have simulated $1:2$ and $1:\infty$, 
with qualitatively similar results; here we state results only for $1:2$.

To simulate the $1:2$ model at $\mu$-values
$\mu_0 < \mu_1 < ... < \mu_{l-1}$, we start with 
$\mu = \mu_0$ and a configuration which is a cycle with $6$ edges. 
The end configuration in the simulation of $\mu_i$ is taken as the starting 
configuration in the simulation of $\mu_{i+1}$. 
The basic Monte Carlo step is: pick at random either an edge $e_1$ or a pair 
of intersecting edges $e_2,e_3$, and then replace $e_1$ with a pair of 
intersecting edges $e_2',e_3'$, or replace $e_2,e_3$ with an edge $e_1'$, 
with probabilities determined by $m_{\mu_i}$, such that the resulting 
configuration is legal (i.e. is a self-avoiding closed loop). 

For each measurement $meas$ we use a standard autocorrelation function to 
find a ``mixing time'' at each $\mu_i$, which we define as the smallest value 
of $t$ such that the autocorrelation
$$A(t) := \frac{1}{(m-t)\sigma^2}\sum_{i=1}^{m-t}
(meas(C_i)-\mu)\cdot(meas(C_{i+t})-\mu)
\eqno{(3)}$$
falls below zero. Here $C_1,...,C_m$ is the Monte Carlo chain corresponding 
to the simulation of $\mu_i$, and $\mu$ and $\sigma^2$ are the
(sample) average and 
variance of $meas$ over 
that chain. We found that for each of our measurements $meas$ (described below), 
our simulations of each $\mu_i$ were on average at least 20 times as long as the corresponding 
mixing times. We therefore believe our simulations are in equilibrium at each $\mu_i$. 
We obtain error bars from the Student's $t$-distribution by running $200$ independent copies
of the simulation. 
 
We make the following measurements to detect the spontaneous symmetry 
breaking and layering which may occur at large $\mu$. 
We first consider a correlation measurement ${corr}(C)$: 
choose a random edge in $C$ and 
define $corr(C)$ as the proportion of edges in $C$ which are parallel to it. 
Since the model is isotropic we expect $corr(\mu)$ to be identically $1/3$ 
for small $\mu$ in the infinite volume limit. 

To detect bulk-sized layers, that is, layers 
proportional to the size of the system, we define 
$lay(C)$ as the size of the largest $80\%$ perfect square ``layer'' centered 
at the origin in $C$, divided by the system size. (Perfect means no bends.)
We expect that for small $\mu$, $lay$ is identically zero in the infinite volume limit. 
Note that the choice of $80\%$ is rather arbitrary; any percentage significantly
above $33\%$ should detect bulk layers.

\begin{figure}\label{fig3}
\centerline{\epsfxsize=6.5truein \epsfbox{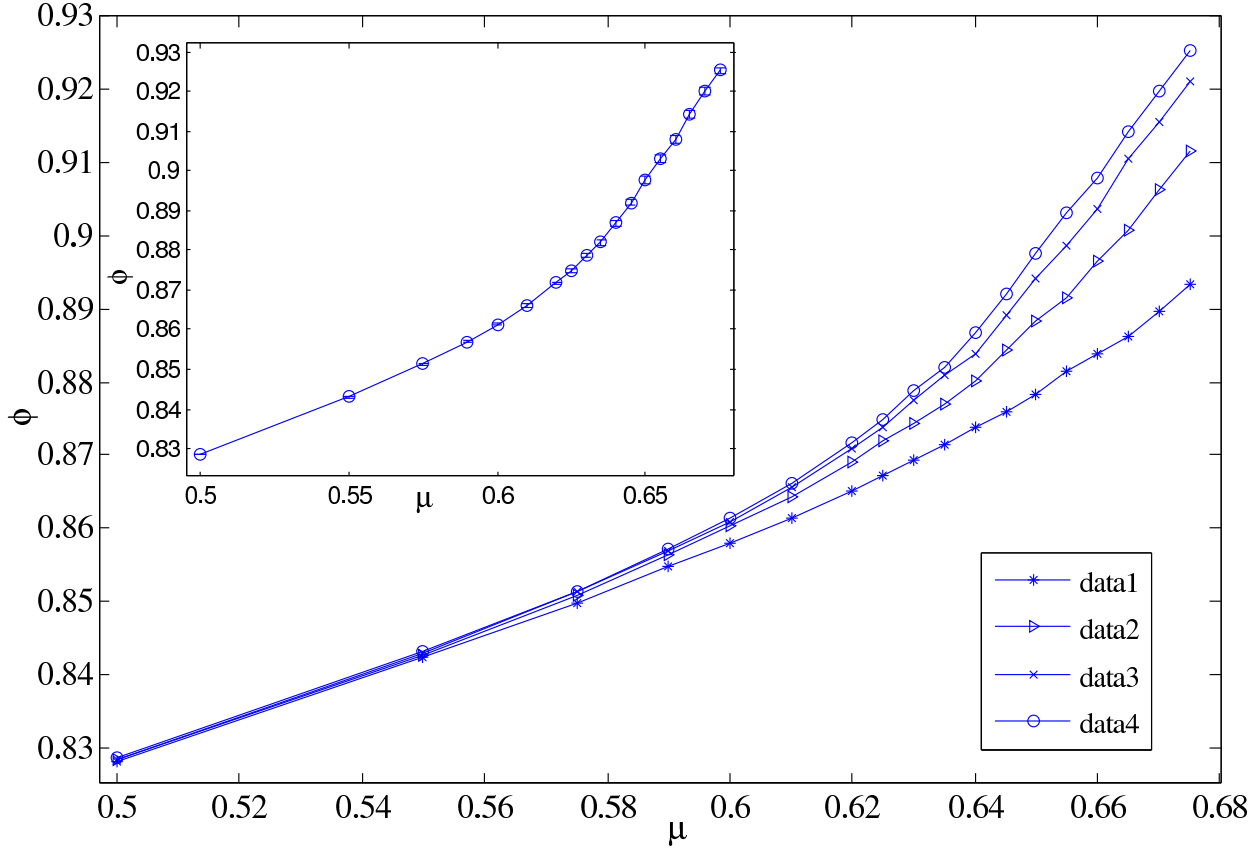}}
\caption{Volume fraction vs. mu, for volumes $40^2$ (data1) through $100^2$ (data4).
Error bars in the inset figure represent 95$\%$ confidence intervals for
volume $100^2$, and are mostly smaller than the data circles.}
\end{figure}

The data gives strong evidence of a phase transition, namely that 
in the infinite volume limit $corr(\mu)$ is identically $1/3$ for
$\mu < \mu^*$ and $corr(\mu) > 1/3$ for $\mu > \mu^*$, where $\mu^* \approx 0.63$. 
See Figure 1a, with error bars in the inset. The data in Figure 1 are performed on
systems of volume: $40^2=1600,\ 60^2=3600,\ 80^2=6400$ and
$100^2=10,000$.  

As further detail of the nature of the ordered phase signalled by $corr$, 
the data displayed in Figure 1b, computed in the same family of simulations,
suggests that in the infinite volume limit $lay$ is 
identically zero below $\mu^*$, but positive above $\mu^*$, showing 
the emergence of bulk layers above $\mu^*$. The transition can be 
seen in configuration
snapshots; see Figure 2. We also measure volume fraction, $\phi(\mu)$. 
The data in Figure 3 suggests that in the infinite volume limit a
discontinuity develops in the slope of $\phi(\mu)$ at $\mu^*$, in
accord with the other evidence of a transition.

\section{Summary}

We have introduced and simulated a toy model for the
folding of progressively confined stiff sheets. 
We find that bulk folding emerges at a sharp volume fraction 
as the material is compacted, just as bulk solids form at 
a sharp volume fraction in the freezing transition of equilibrium fluids. 
This analogy has previously been used to model the behavior of other 
types of soft matter, in particular colloids [9] and the random close 
packing of granular matter [10,11].

We note an old model due to Flory [12] which has been applied to
crumpled materials; see [13] and references therein. In that two
dimensional lattice model one varies the temperature with volume
fraction fixed at one. There is general agreement that it exhibits a
phase transition, perhaps representing the melting of a dense ordered
phase. This melting transition is usually characterized as continuous
(second order), though this is sometimes disputed [13].

The phase transition we find in our model should be experimentally verifiable,
in both compacted stiff sheets and in two dimensionally confined
compacted stiff wires.

\section*{Acknowledgements} 

The authors gratefully acknowledge useful
discussions with W.D. McCormick, N. Menon and H.L. Swinney on the
experimental possibilities of crumpled materials, and financial
support from NSF Grant DMS-0700120 and Paris Tech (E.S.P.C.I.)



\begin{thebibliography}{99}

\bibitem{1} T.A. Witten, 
Stress focusing in elastic sheets, 
Rev. Mod. Phys., 79 (2007), 643-675. 

\bibitem{2} K. Matan, R.B. Williams, T.A. Witten and S.R. Nagel, 
Crumpling a thin sheet, 
Phys. Rev. Let.  88, No. 7 (2002) 076101.

\bibitem{3} A.S. Balankin and O.S. Huerta,
Entropic rigidity of a crumpling network in a randomly folded thin sheet, 
Phys. Rev. E 77 (2008) 051124.

\bibitem{4} L. Bou\'e and E. Katzav, 
Folding of flexible rods confined in 2D space, 
Europhys. Lett 80 (2007) 54002.

\bibitem{5} Y. Kantor,
Properties of tethered surfaces,
in ``Statistical Mechanics of Membranes and Surfaces - Proceedings of
the Fifth Jerusalem Winter School for Theoretical Physics'', ed. by
D. R. Nelson, T. Piran and S. Weinberg, World Scientific,
Singapore, (1989) pp. 115-136.

\bibitem{6} A. Sultan and A. Boudaoud, 
Statistics of crumpled paper,
Phys. Rev. Let. 96 (2006) 136103.

\bibitem{7} Y.C.Lin, Y.W. Lin and T.M. Hong,
Crumpling wires in two dimensions,
Phys. Rev. E 78 (2008) 067101. 

\bibitem{8}  E. Katzav, M. Adda-Bedia and A. Boudaoud,
A statistical approach to close packing of elastic rods and to DNA
packaging in viral capsids,
Proc. Nat. Acad. Sci. 103 (2006) 18900-18904.

\bibitem{9}  M.A. Rutgers, J.H. Dunsmuir, J.-Z. Xue, W.B. Russel 
and P.M. Chaikin, 
Measurement of the hard-sphere equation of state using screened charged 
polystyrene colloids, Phys. Rev. B 53 (1996) 5043-5046.  

\bibitem{10} C. Radin, 
Random close packing of granular matter,
J. Stat. Phys. 131 (2008), 567-573.

\bibitem{11} D. Aristoff and C. Radin, 
Random close packing in a granular model,
arXiv:0909.2608.

\bibitem{12} P.J. Flory, 
Statistical thermodynamics of semi-flexible chain molecules,
Proc. R. Soc. London, Ser. A 234 (1956) 60-73.

\bibitem{13} J.L. Jacobsen and J. Kondev,
Conformal field theory of the Flory model of polymer bending,
Phys. Rev. E 69 (2004) 066108.

\end{thebibliography}
\end{document}